\input phyzzx
\input epsf

\def\scrip{{\cal I^+}}
\def\scrim{{\cal I^-}}
\def\tR{{\tilde R}}
\def\ty{{\tilde y}}
\def\tU{{\tilde U}}
\def\tV{{\tilde V}}

\rightline{UATP-98/01}
\rightline{February 1998}
\vskip 0.1in
\centerline{\seventeenbf Particle Creation in the Marginally Bound,}
\centerline{\seventeenbf Self Similar Collapse of Inhomogeneous Dust}
\vskip 0.25in
\centerline{\caps S. Barve, T. P. Singh\footnote{\dagger}{Email:
sukkoo@relativity.tifr.res.in, tpsingh@tifrc3.tifr.res.in}}
\centerline{\it Tata Institute of Fundamental Research}
\centerline{\it Homi Bhabha Road, Bombay 400-005, India}
\vskip 0.1in
\centerline{\caps Cenalo Vaz\footnote{\dagger\dagger}{Email:
cvaz@mozart.si.ualg.pt; (to whom correspondence should be addressed)}}
\centerline{\it Unidade de Ci\^encias Exactas e Humanas}
\centerline{\it Universidade do Algarve}
\centerline{\it Campus de Gambelas, P-8000 Faro, Portugal}
\vskip 0.1in
\centerline{and}
\vskip 0.1in
\centerline{\caps Louis Witten\footnote{\dagger\dagger\dagger}{Email:
witten@physunc.phy.uc.edu}}
\centerline{\it Department of Physics}
\centerline{\it University of Cincinnati}
\centerline{\it Cincinnati, OH 45221-0011, U.S.A.}
\vskip 0.25in
\centerline{\bf \caps Abstract}
\vskip 0.1in

\noindent We consider the evaporation of the (shell focusing) naked singularity
formed during the self-similar collapse of a marginally bound inhomogeneous dust cloud,
in the geometric optics approximation. We show that, neglecting the back reaction of
the spacetime, the radiation on $\scrip$ tends to infinity as the Cauchy Horizon is
approached. Two consequences can be expected from this result: (a) that the back
reaction of spacetime will be large and eventually halt the formation of a naked
singularity thus preserving the Cosmic Censorship Hypothesis and (b) matter attempting
to collapse into a naked singularity will radiate away energy at an intense rate, thereby
possibly providing experimental signatures of quantum effects in curved spacetimes.
\vfill
\eject

It is expected that very massive stars will undergo continual gravitational
collapse. While the singularity theorems of Geroch, Hawking and Penrose establish
that, under fairly general conditions, such a gravitational collapse will result
in the formation of a singularity, they do not, by themselves, indicate
whether the singularity will be hidden behind an event horizon or whether it
will be visible to an outside observer. If the singularity is hidden the
collapse ends in a black hole, and if it is visible, the collapse ends in a
naked singularity. It is therefore an open problem in classical general relativity
as to whether gravitational collapse ends in a black hole or in a naked
singularity. Naked singularities have, so far, been considered
undesirable and this has produced the so-called ``cosmic censorship'' hypothesis
(CCH).${}^{[1]}$ The CCH roughly states that the singularities arising in gravitational
collapse of ``reasonable'' forms of matter with ``reasonable'' initial conditions
are always hidden behind horizons and are not visible. Yet, attempts to prove this
hypothesis on the classical level have been unsuccessful. On the contrary, studies
of classical models of gravitational collapse have shown that both black holes
and naked singularities can arise in collapse, depending on initial conditions.${}^{[2]}$
If naked singularities do arise generically in collapse, they very possibly will
have important observational astrophysical consequences.

Most studies of gravitational collapse assume spherical symmetry and even in
this simplest of cases our understanding of the outcome of collapse is
incomplete. The first model to be studied was the collapse of a homogeneous
dust sphere (the Oppenheimer Snyder model).${}^{[3]}$ This results in the formation of a black
hole and most of our understanding of how a black hole forms in collapse, indeed,
the censorship hypothesis itself, is motivated by this model. Later, the spherical collapse
of inhomogeneous dust was investigated. This system is described by an exact solution of
Einstein equations, which was given independently by Tolman and by Bondi.${}^{[4]}$ The nature
of the resulting singularities has been investigated by various authors${}^{[5]}$ and it has been
found that while some of the initial density and velocity distributions lead to black hole
formation, other distributions result in the formation of naked singularities. There is a
smooth transition from one phase to the other, and the solution of Oppenheimer and Snyder is
a very special case of a general inhomogeneous class.${}^{[6]}$

Another well-studied system is the spherical collapse of null dust, describing an exact solution
of Einstein equations, the Vaidya spacetime. Once again, it has been found that both Black Holes
and Naked Singularities result in this collapse, depending on the rate of infall of the null
dust.${}^{[7]}$

Unfortunately, analyses of models of collapse with more realistic equations of state are
hindered by the paucity of physically reasonable exact solutions. The collapse of
a self-similar perfect fluid was investigated numerically by Ori and Piran${}^{[8]}$ who found
generic naked singularity solutions. These solutions, on their own merit, represent a serious
violation of cosmic censorship.  Further, numerical studies of collapsing scalar fields${}^{[9]}$
indicate that the field disperses entirely without forming a singularity, in the domain of weak
gravitational coupling. In the limit when the coupling is very strong, all the mass collapses
to form a black hole, but, in the intermediate regime, part of the mass collapses to form a
black hole and rest of it disperses. The transition point between dispersive and singular
behavior is a naked singularity. The collapsed mass also shows a power-law dependence
on the difference of the coupling parameter from its critical value, and the power-law index
is independent of initial conditions. Similar behavior has also been found in the collapse of
other forms of matter.${}^{[10]}$

It is likely of course that the initial conditions that lead to the classical collapse of
matter into naked singularities are not acceptable quantum conditions. It is also likely
that quantum effects, for example particle production in the presence of the strong
gravitational fields involved toward the final stages, dominate the evolution at late
times and prevent their formation. In this article we argue in favor of the latter. We will
consider below a spherically symmetric model of inhomogeneous, marginally bound self
similar dust collapse that arises as a special case of the general Tolman-Bondi collapse
problem, analyze the causal structure of the spacetime and calculate the leading order
contribution to the radiated power in the geometric optics approximation. We find that the
radiated power diverges as the inverse square of the retarded distance from the
Cauchy Horizon.

More specifically, the model is a solution of Einstein's equations with matter described by
the stress energy tensor
$$T_{\mu\nu}~~ =~~ \epsilon(t,r) \delta^0_\mu \delta^0_\nu.\eqno(1)$$
The metric is well known and given in comoving coordinates by
$$ds^2~~ =~~ dt^2~ -~ \tR^{'2}(t,r)  dr^2~ -~ \tR^2(t,r) d\Omega^2 \eqno(2)$$
where the dust cloud is thought of as made up of concentric shells, each labeled by $r$.
$\tR'(t,r)$ is the derivative of $\tR(t,r)$ with respect to $r$ and $\tR(t,r)$ is the physical
radius (the area of a shell labelled by $r$ is $4\pi R^2(t,r)$) obeying, in the particular
case of the marginally bound self similar collapse,
$$\tR(t,r)~~ =~~ r \left[1~ -~ {{3\sqrt{\lambda}} \over 2} {t \over r} \right]^{2/3}.
\eqno(3)$$
The physical radius is seen to depend on one parameter, $\lambda$, (the ``mass parameter'').
This parameter determines the total mass, $M(r)$, lying within the shell labeled by $r$ as
$2GM(r) = \lambda r$. The total mass of the dust is therefore $2GM = \kappa = \lambda r_o$
where $r_o$ labels the outer boundary of the cloud. Now it can be shown that $\tR(t,r) = 0$ is
a curvature singularity. This means that the singularity curves are
$t_o(r) = 2r/ (3\sqrt{\lambda})$, so that the last shell becomes singular at the time
$t_o = 2/3 \sqrt{r_o^3/ \kappa}$.

Beyond $r=r_o$ one has the Schwarzschild spacetime
$$ds^2~~ =~~ \left(1 - {{\kappa} \over R}\right) dT^2~ -~ \left( 1 -
{{\kappa} \over R}\right)^{-1} dR^2~ -~ R^2 d\Omega^2 \eqno(4)$$
and the Tolman Bondi spacetime in (2) must be matched to (4) at the boundary. This means
that the two metrics must agree on the three dimensional hypersurface  traced out by the
outer boundary $r=r_o$. Comparing the angular coordinates, one concludes that
$$\tR(t,r_o)~~ =~~ R_o(t)$$
Therefore, requiring that the first fundamental forms agree on the hypersurface traced out
by the collapsing outer boundary one has
$$\left(1 - {{\kappa} \over {\tR_o(t)}}\right) \left[{{dT} \over {dt}}
\right]^2~~ =~~ 1~ +~ \left(1 - {{\kappa} \over {\tR_o(t)}}\right)^{-1}
\left[{{d\tR_o(t)} \over {dt}} \right]^2 \eqno(5)$$
where $\tR_o(t)= \tR(r_o,t)$. Using ${\dot \tR}(r_o,t)=-{\sqrt{\kappa/\tR}}$
one finds that (5) gives
$$T_o(t)~~ =~~ \int dt \left(1 - {{\kappa} \over {\tR_o(t)}}\right)^{-1}~~
=~~ -~ \int d\tR_o(t) {\sqrt{{\tR_o(t)} \over {\kappa}}} \left(1 - {{\kappa}
\over {\tR_o(t)}}\right)^{-1} \eqno(6)$$
which integral may be solved to give
$$\eqalign{T_o(t)~~ &=~~ -~ 2{\sqrt{\kappa\tR_o}}~ -~ {2 \over 3} \tR_o {\sqrt{
{\tR_o} \over {\kappa}}}~ +~ \kappa \ln |{{{\sqrt{\tR_o}} + {\sqrt{\kappa}}} \over
{{\sqrt{\tR_o}} - {\sqrt{\kappa}}}}|\cr &=~~ t~ -~ {2 \over {3\sqrt
{\kappa}}}r_o^{3/2}~ -~ 2 \sqrt{{\kappa \tR_o}}~ +~ \kappa \ln |{{{\sqrt{\tR_o}} +
{\sqrt{\kappa}}} \over {{\sqrt{\tR_o}} - {\sqrt{\kappa}}}}|\cr R_o(t)~~ &=~~ r_o \left[
1 - a {t \over r_o} \right]^{2/3}\cr} \eqno(7)$$
where we have set $a = 3\sqrt{\lambda}/2$. One can then show that the second
fundamental forms agree by the relations in (7).

For the marginally bound, self-similar collapse under consideration, it is relatively
simple to find null coordinates for this system. Consider the effective two dimensional
metric
$$ds^2~~ =~~ dt^2~ -~ \tR^{'2}(t,r) dr^2 \eqno(8)$$
and change variables to $z,x$ where $z~ =~ \ln r$, $x=t/r$. This gives
$$\eqalign{ds^2~~ &=~~ r^2 \left[dx^2~ +~ 2xdxdz~ +~ (x^2~ -\tR^{'2}(x)) dz^2\right]\cr
&=~~ r^2 (x^2 - \tR^{'2}) (d\tau^2~ -~ d\chi^2)\cr}\eqno(9)$$
where
$$\eqalign{\tau~~ =~~ z~ +~ {1 \over 2} (I_- + I_+)\cr \chi~~ =~~ {1 \over 2}
(I_- - I_+)\cr}\eqno(10)$$
in terms of
$$I_\pm (x)~~ =~~ \int {{dx} \over {x\pm\tR'}}\eqno(11)$$
We would like to choose null coordinates such that in the limit as $\lambda
\rightarrow 0$ these reduce to the standard null coordinates in Minkowski
space. Such coordinates are given by
$$\eqalign{u~~ &=~~ \left\{\matrix{+r e^{I_-}&~~~ x-\tR' > 0\cr-r e^{I-}&~~~ x-
\tR' < 0}\right.\cr v~~ &=~~ \left\{\matrix{+r e^{I_+}&~~~ x+\tR' > 0\cr-r e^{I_+}
&~~~ x+\tR'<0}\right.\cr}\eqno(12)$$

\noindent To further analyze the causal structure, it is now convenient to go over to the
variable $y$ defined by $y = \sqrt{\tR/r}$. In terms of $y$, the integrals $I_\pm$ can be written
as
$$I_\pm~~ =~~ 9 \int {{y^3 dy} \over {3y^4~ \mp~ ay^3~ -~ 3y~ \mp~
2a}} \eqno(13)$$
and the coordinates (12) become
$$\eqalign{u~~ &=~~ \left\{\matrix{+r e^{I_-}&~~~ f_-(y) < 0\cr-r e^{I-}&~~~ f_-(y)
> 0}\right.\cr v~~ &=~~ \left\{\matrix{+r e^{I_+}&~~~ f_+(y) < 0\cr-r e^{I_+}
&~~~ f_+(y) > 0}\right.\cr}\eqno(14)$$
where
$$f_\pm (y)~~ =~~ 3y^4~ \mp~ ay^3~ -~ 3y~ \mp~ 2a. \eqno(15)$$
Let $\alpha^\pm_i$ be the roots of $f_\pm(y)$, for $i~ \epsilon~ \{1,2,3,4\}$. As $f_\pm$
are both real, they admit either 0, 2, or 4 real roots. The integrals can now be put in the form
$$I_\pm~~ =~~ 3 \int dy \left[\sum_{i=1}^4 {{A^\pm_i} \over {(y-\alpha^\pm_i)}}\right]
\eqno(16)$$
where the $A^\pm_i$ are constants related to the coefficients of $f_\pm(y)$ and their
roots by,
$$A_i^\pm~~ =~~ {{\alpha_i^{\pm 3}} \over {f'_\pm(\alpha_i^\pm)}} \eqno(17)$$
In particular, the $A^\pm_i$ satisfy $\sum_i A^\pm_i = 1$. If all the roots are real,
the solution is explicitly given by
$$\eqalign{u(y)~~ &=~~ \pm~ r \prod_{i=1}^4 |y~ -~ \alpha^-_i|^{3A^-_i}\cr
v(y)~~ &=~~ \pm~ r \prod_{i=1}^4 |y~ -~ \alpha^+_i|^{3A^+_i}\cr}\eqno(18)$$
We will now consider the case in which there are two real roots and a conjugate
pair of complex roots. As we will shortly show at least two real roots (possibly degenerate)
are required for the existence of a globally naked singularity at the origin so we do not
consider the case when all the roots are complex even though it may be carried out in
the same spirit. Let us order the roots so that the first two, $\alpha_{1,2}$, are a
complex conjugate pair and $\alpha_{3,4}$ are real. From (17) it follows that
$A_{1,2}$ is also a complex pair whereas $A_{3,4}$ are real. Then the integrals
are of the form
$$I~~ =~~ 3 \int dy \left[\sum_{i=1}^4 {{A_i} \over {(y-\alpha_i)}}\right]~~ =~~
3 \left[A \ln (y - \alpha)~ +~ A^* \ln (y - \alpha^*)~ +~ \sum_{i=3,4} A_i \ln |
y - \alpha_i|\right]\eqno(19)$$
where $\alpha, \alpha^*$ are the complex roots and $A,A^*$ are the (complex)
coefficients. Putting
$$A~~ =~~ |A| e^{i\phi},~~~~~~~~~~ y-\alpha~~ =~~ |y-\alpha| e^{i\xi}\eqno(20)$$
so that the $u,v$ coordinates have the explicit (and formal) solution
$$\eqalign{u(y)~~ &=~~ \pm r|y-\alpha^-|^{6|A^-|\cos \phi^-}e^{-6|A^-|\xi^-\sin\phi^-}
\Pi_{i=3,4} |y-\alpha^-_i|^{3A^-_i}\cr v(y)~~ &=~~ \pm r|y-\alpha^+|^{6|A^+|\cos\phi^+}
e^{-6|A^+|\xi^+\sin\phi^+} \Pi_{i=3,4} |y-\alpha^+_i|^{3A^+_i}\cr}\eqno(21)$$
Consider the center ($r=0$) at early times, $t<0$. Then, because $y = (1-at/r)^{1/3}
\rightarrow \infty$, (18) gives (when all roots are real)
$$\eqalign{u~~ &\rightarrow~~ -~ r |y|^{3\sum_i A^-_i}~~ =~~
-r (1~ -~ a{t\over r})~~ \rightarrow at\cr v~~ &\rightarrow~~ -~ r |y|^{3\sum_i A^+_i}~~
=~~ -r (1~ -~ a{t\over r})~~ \rightarrow at \cr}\eqno(22)$$
This line is therefore given by $u = v$. When two of the roots are conjugate complex the
line is still $u=v$ as we now show. Note that
$$\xi~~ =~~ \tan^{-1} \left({{{\rm Im} (-\alpha)} \over {{{\rm Re}(y-\alpha)}}}\right)$$
($y$ is real), so that as $y\rightarrow \infty$, $\xi \rightarrow 0$. Then
clearly
$$\eqalign{u~~ &\rightarrow~~ -~ r |y|^{3(2{\rm Re} A^- + A^-_3 + A^-_4)}\cr v~~ &
\rightarrow~~ -~ r |y|^{3(2{\rm Re} A^+ + A^+_3 + A^+_4)}\cr}\eqno(23)$$
but since $\sum_i A^\pm_i = 1$, we have the same result as before.

The general solutions in equations (18, 21) are useful to analyze another limit, namely
the singularity at $r \rightarrow at$. This means that $y \rightarrow 0$. Now when
$y\rightarrow 0$, $f_-(y)>0 $ and $f_+(y)<0$. Then we see that (if all roots are real)
$$\eqalign{u~~ &=~~ -~ r \prod |\alpha^-_i |^{3A^-_i}\cr v~~ &=~~ r \prod |\alpha^+_i
|^{3A^+_i}\cr}\eqno(24)$$
and, in particular,
$${v\over u}~~ =~~ -~ c~~ =~~ -~ \prod_i {{|\alpha^+_i |^{3A^+_i}} \over
{|\alpha^-_i |^{3A^-_i}}} \eqno(25)$$
which is a negative constant, in general $\neq -1$. The singularity is therefore {\it
spacelike} until the last shell, $r=r_o$, collapses at $t=t_o = r_o/a$. The case
of a pair of conjugate complex roots trivially gives the same result. Beyond this point
the singularity will be spacelike because it is just the Schwarzschild singularity in
the exterior region. The behavior of the origin, $r=0, t=0$, is peculiar. It is
the meeting point between two lines $u=v$ and $u=-cv$ and it's nakedness (coveredness)
is far from clear. However, if a null ray originating at this point reaches the
boundary at Kruskal coordinate $U < 0$ in the Schwarzschild region, it will reach
$\scrip$ and then the origin will be globally naked.

We will be interested in the earliest null ray leaving the singularity and reaching
$\scrip$ (the Cauchy Horizon) as well as the earliest null ray that strikes the singularity
from $\scrim$. These rays can be expected to intersect the first singular
shell at $r=0, t=0$, so it is natural to carefully examine the null rays passing through this
point. The origin, being the intersection of the lines $u=v$ and $v= -cu$ ($c \neq 1$ in general),
corresponds to the point $u=0=v$. Now any null ray traveling toward $\scrip$ with $u=0$ must
have either $r=0$ or $I_- \rightarrow -\infty$. Therefore, when $r \neq 0$, such a ray is possible
if and only if $y=\alpha^-_k$ for some {\it real} root, $\alpha^-_k$ of the polynomial $f_-(y)$. Indeed
such a root may not exist, in which case the singularity is not naked as no null rays can emanate
from it. If a real root exists however, at least one null ray leaves this point and reaches the
boundary. The existence of real roots of the polynomials $f_-(y)$ is therefore a necessary condition
for the nakedness of the origin. This places a constraint on the possible value of the constant $a$
in the mass function. One finds that real roots exist provided that $a < a_c \sim 0.638$.${}^{[11]}$
Each root corresponds to a null ray emanating from $u=0=v$ and there are at least two of them, if any
at all. Because $y = \alpha_i$ implies that $t = r(1-\alpha_i^3)/a$, we choose the largest real
root of $f_-(y)$ as the one that gives the earliest null ray emanating from $u=0=v$ and call it
$\alpha_-$. Thus, $y=\alpha_-$ is the Cauchy horizon.

A similar reasoning can now be given for the incoming rays passing through $u=0=v$. Again any
ray with $v=0$ for $r\neq 0$ must have $I_+ \rightarrow -\infty$, which is possible only
if $y=\alpha^+_k$ for some {\it real} root, $\alpha^+_k$ of the polynomial $f_+(y)$. Now,
$f_+(y)$ admits two real roots, one unphysical (negative) and one positive.  Again,
call the (positive) physical root $\alpha_+$.

What we have described above is pictured in the Penrose diagram of figure I.
\vskip 0.2in

\centerline{\epsfxsize=2.5in \epsfbox{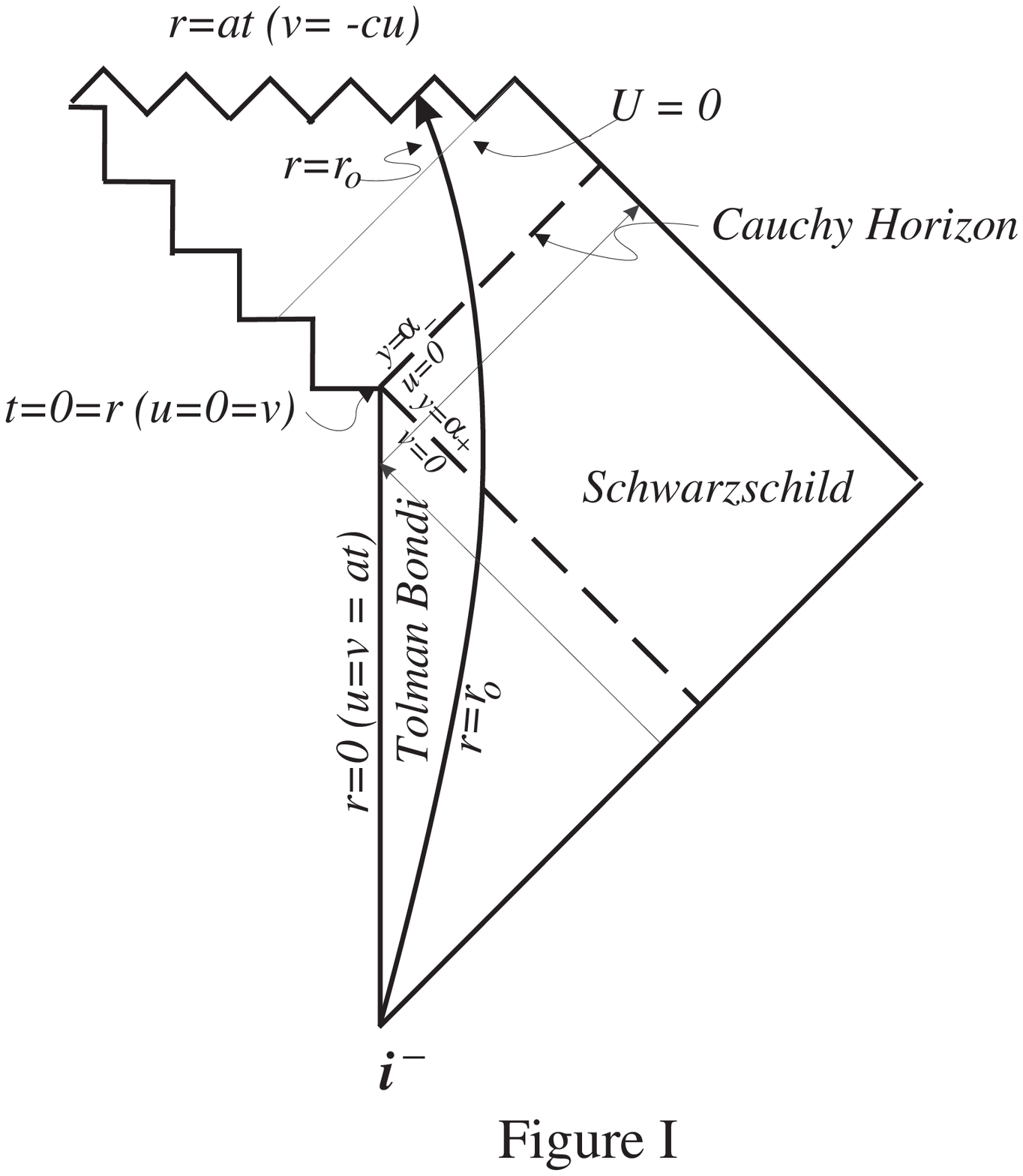}}
\vskip 0.2in

\noindent We will henceforth consider rays in the neighborhood of the lines given by $y = \alpha_-$
for outgoing rays and $y = \alpha_+$ for incoming rays. The precise values of $\alpha_\pm$ in terms
of the mass parameter will not interest us for this work but we will Taylor expand about these two
values, considering $y_\pm = \ty_\pm + \alpha_\pm$.

Returning to (7), one can rewrite the Schwarzschild radial coordinate and time on the boundary as
follows
$$\eqalign{R_o(y)~~ &=~~ r_o y^2\cr T_o(y)~~ &=~~ -~ {{r_o} \over a} y^3~ -~  {4 \over 3} a r_o y~
-~ {4 \over 9} a^2 r_o \ln |{{3y/2a - 1} \over {3y/2a + 1}}| \cr}\eqno(26)$$
Therefore, the Eddington-Finkelstein null coordinates on the boundary, $\tU_o(y) = T_o(y)-R_{o*}(y)$,
$\tV_o(y) = T_o(y) + R_{o*}(y)$, (where $R_{o*}$ is the tortoise coordinate) take the form
$$\eqalign{\tU_o(y)~~ &=~~ -~ {{r_o} \over a} y^3~ -~  {4 \over 3} a r_o y~ -~ r_o y^2 ~ -~
{8\over 9} a^2 r_o \ln |3y/2a - 1|\cr \tV_o(y)~~ &=~~ -~ {{r_o} \over a} y^3~ -~  {4 \over 3}
a r_o y~ +~ r_o y^2 ~ +~ {8\over 9} a^2 r_o \ln |3y/2a + 1|\cr}\eqno(27)$$
It is now clear that the earliest null outgoing ray, $u=0$, from the origin (the Cauchy Horizon)
within the cloud strikes the boundary at $y=\alpha_-$ and translates into the null outgoing ray
$$\tU_o^{(0)}~~ =~~ -~ {{r_o} \over a} \alpha_-^3~ -~  {4 \over 3} a r_o \alpha_-~ -~ r_o
\alpha_-^2 ~ -~ {8\over 9} a^2 r_o \ln |3 \alpha_-/2a - 1|\eqno(28)$$
which is never infinite ($2a/3$ is not a root of $f_-(y)$). This null ray corresponds to a finite
value of $\tU$ and will therefore reach $\scrip$, so the existence of real roots of $f_-(y)$
turns out to be not just necessary, but a sufficient condition for the origin to be globally naked.
The same argument applies to the infalling ray(s): the earliest null ray to pass
through the origin is the ray corresponding to the value $y=\alpha_+$, or
$$\tV_o^{(0)}~~ =~~ -~ {{r_o} \over a} \alpha_+^3~ -~  {4 \over 3} a r_o \alpha_+~ +~ r_o
\alpha_+^2 ~ +~ {8\over 9} a^2 r_o \ln |3 \alpha_+/2a + 1|\eqno(29)$$
and, again, since $-2a/3$ is not a root of $f_+(y)$, $\tV$ is not infinitely negative and such a
ray will have come from $\scrim$. Thus, the existence of a positive real root of $f_+(y)$ is
sufficient to ensure that at least one infalling ray from $\scrim$ will intersect the origin.

The next question we must address is the relationship between the $\tU, \tV$ coordinates in the
exterior and the $u,v$ coordinates (equations (18, 21)) on the boundary. This is difficult to do in
general, but if we confine our study to rays that are ``close'' to $u=0$ and $v=0$ we can arrive at
some conclusion regarding the quantum radiation on $\scrip$ near the Cauchy horizon. ``Close''
will be taken to mean linearizations about $y=\alpha_\pm$ respectively for incoming rays and
outgoing rays.

First consider outgoing rays. For $y \sim \alpha_-$, define $y=\ty+\alpha_-$ and find that for small
$\ty$
$$I_-~~ \sim~~ \gamma_- \ln \ty~~ +~~ {\cal O}(y)\eqno(30)$$
where
$$\gamma_-~~ =~~ {{3\alpha_-^3} \over {f'_-(\alpha_-)}}\eqno(31)$$
giving
$$u~~ =~~ -r|\ty|^{\gamma_-}~~ \rightarrow~~ y-\alpha_-~~ =~~ \left(- {u \over r} \right)^{1/
\gamma_-}\eqno(32)$$
Therefore in terms of $u$ (on the boundary) we can write $\tU$ as follows
$$\tU~~ \sim~~ \tU^{(0)}(\alpha_-)~ +~ \Gamma_-(\alpha_-) (y-\alpha_-) ~~ =~~ \tU^{(0)}(
\alpha_-)~ +~ \Gamma_-(\alpha_-) \left(-{{u} \over {r_o}} \right)^{1/\gamma_-}\eqno(33)$$
where
$$\Gamma_-~~ =~~ -~9 {{r_o \alpha_-^3} \over {a (3\alpha_- -2a)}}~~ <~~ 0~~ {\rm when}~~ a < a_c
\eqno(34)$$
Figure II is a plot of $\Gamma_-$ as a function of $a$ for $a<a_c$.
\vskip 0.1in
\centerline{\epsfxsize=2.5in \epsfbox{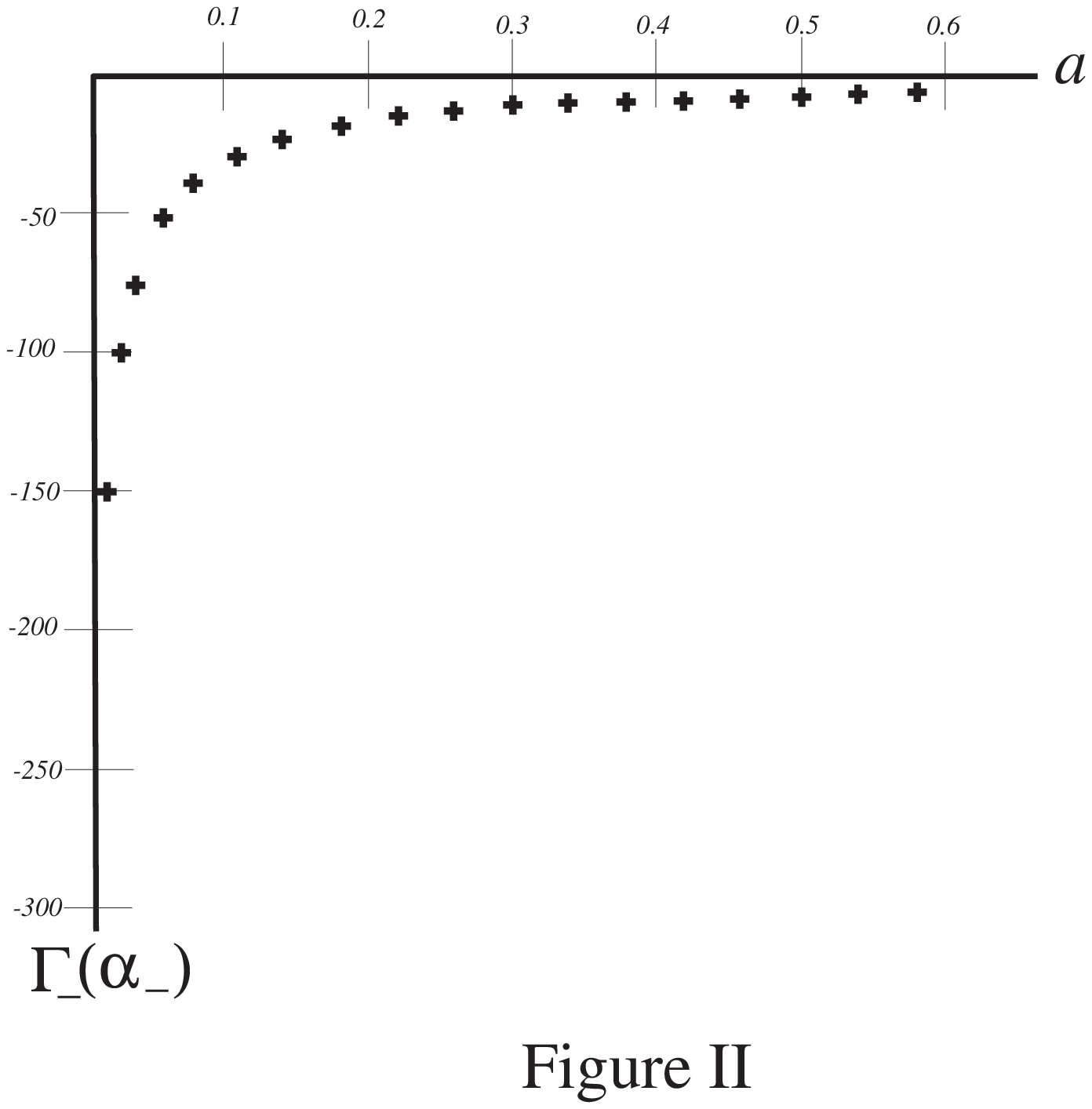}}
\vskip 0.1in
\noindent Likewise, for incoming rays,
put $y = \ty+\alpha_+$ and find that
$$I_+~~ =~~ \gamma_+ \ln \ty~~ +~~ {\cal O}(y)\eqno(35)$$
where
$$\gamma_+~~ =~~ {{3\alpha_+^3} \over {f'_+(\alpha_+)}}\eqno(36)$$
giving
$$v~~ =~~ -r|\ty|^{\gamma_+}~~ \rightarrow~~ y-\alpha_+~~ =~~ \left(- {v \over r} \right)^{1/
\gamma_+}\eqno(37)$$
Thus, in terms of $v$ (on the boundary) we can write $\tV$ as follows
$$\tV~~ \sim~~ \tV^{(0)}(\alpha_+)~ +~ \Gamma_+ (\alpha_+)(y-\alpha_+)~~ =~~ \tV^{(0)}
(\alpha_+)~ +~ \Gamma_+ (\alpha_+) \left(-{{v} \over {r_o}}\right)^{1/\gamma_+}\eqno(38)$$
where
$$\Gamma_+~~ =~~ -~9 {{r_o\alpha_+^3} \over {a (3\alpha_+ + 2a)}}~~ <~~ 0~~  {\rm when}~~  a<a_c
\eqno(39)$$
Figure III is a plot of $\Gamma_-$ as a function of $a$ for $a<a_c$.
\vskip 0.1in
\centerline{\epsfxsize=2.5in \epsfbox{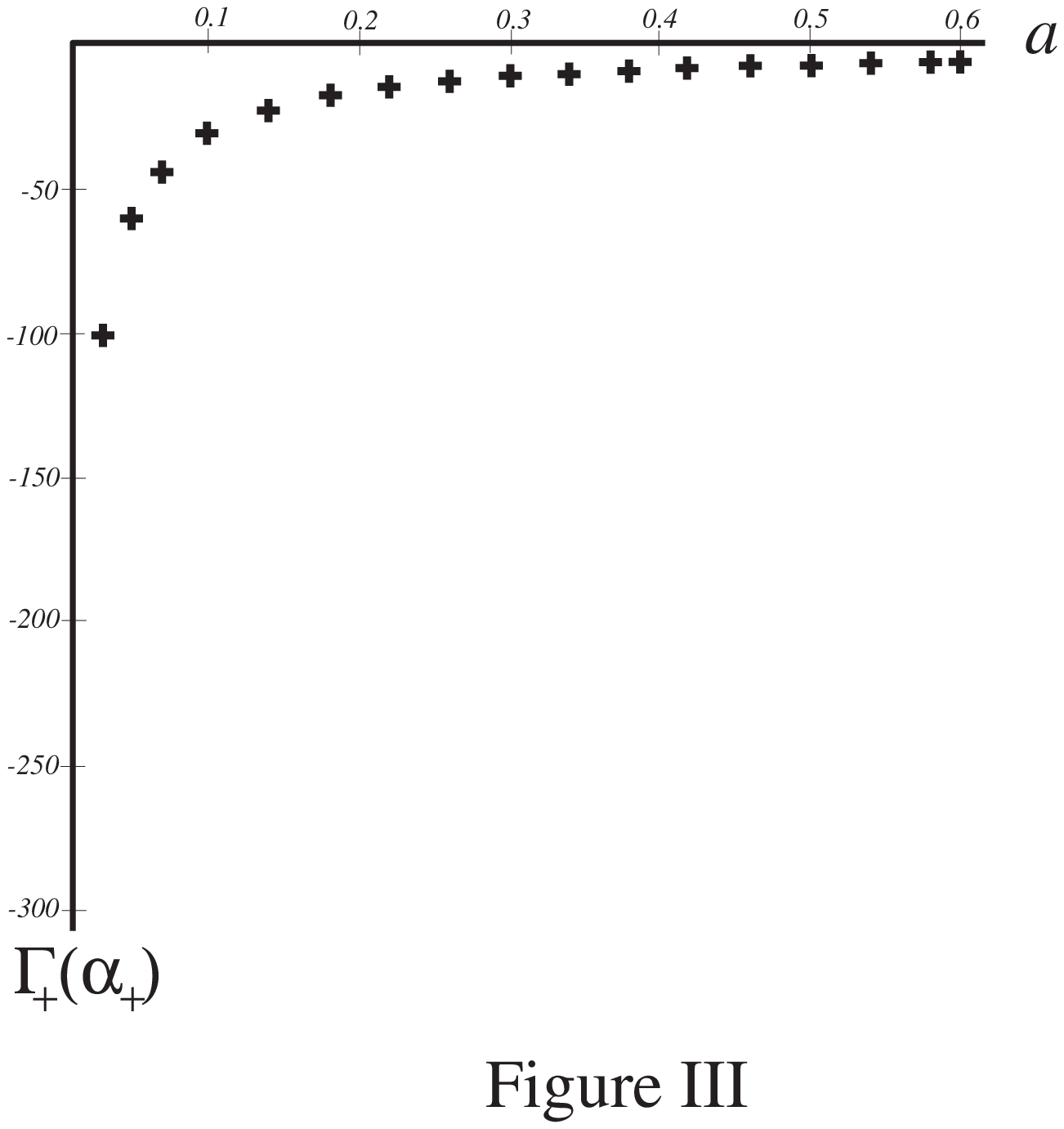}}
\vskip 0.1in
\noindent We are now in a position to compute the radiated power close to the Cauchy horizon in
the geometric optics approximation

The geometric optics approximation is a fairly general way to obtain the power radiated
past $\scrip$ by the lower angular momentum modes in an asymptotically flat spacetime in which
radial null rays define a one to one map between the $\scrim$ and $\scrip$. The method originated
in the work of Moore${}^{[12]}$, Hawking${}^{[13]}$ and DeWitt${}^{[14]}$, was later used
by Fulling and Davies${}^{[15]}$ in a two dimensional model examining the radiation from a moving
mirror and by Ford and Parker${}^{[16]}$ to study some four dimensional collapse models. In their
study, Ford and Parker considered the collapse of a dust cloud leading to the formation of
a shell crossing naked singularity and found that the energy flux of the created scalar
particles remained finite up to the time of formation of the singularity. They also considered the
collapse of charged shells (for which the charge exceeds the mass) leading to naked singularities
and observed that, for these, the flux of created particles is infinite. However, naked
singularities are formed in these models if the proper mass is negative or if Einstein's
equations are not imposed. The model we are considering, on the other hand, is a genuine solution
of Einstein's equations with reasonable matter, yet, as we show, that the result is the same.

Mode
solutions of the free scalar equation $\nabla^2 \phi = 0$ have the general solutions
$$f_{\omega lm}~~ \sim~~ (4\pi\omega)^{-1/2}\left\{\matrix{R^{-1} Y_{lm} e^{-i\omega \tU}\cr R^{-1}
Y_{lm} e^{-i\omega \tV}}\right.\eqno(40)$$
at spatial infinity, $R \rightarrow \infty$, where $\tU$ and $\tV$ are the null coordinates
defined earlier. Consider an expansion of a massless scalar field, $\phi$, in terms of a
complete set of modes
$$\phi~~ =~~ \sum_{lm} \int d\omega [f_{\omega lm} a_{\omega lm}~ +~ f^*_{\omega lm}
a_{\omega lm}^\dagger] \eqno(41)$$
where the $f_{\omega lm}$ are appropriately normalized and reduce to the second of the set in
(40) in the remote past. The `` in'' vacuum is then defined by $a_{\omega lm}|0\rangle = 0$ and
corresponds to the absence of incoming radiation from $\scrim$.

One is interested in the form of $f_{\omega lm}$ in the remote future. An incoming ray, $\tV =$
const., originating at $\scrim$, will pass through the geometry of the spacetime to become an
outgoing null ray that arrives at $\scrip$ at a time $\tU = {\cal F}(\tV)$. Alternatively,
a ray that arrived on $\scrip$ at $\tU$ will have originated on $\scrim$ at $\tV={\cal G}(\tU)$. Thus a
wave packet formed from plane waves $e^{-i\omega \tV}$ becomes, at late times, an
outgoing wave packet formed from plane waves $e^{-i\omega {\cal G}(\tU)}$. It is necessary, therefore,
to consider a solution of the massless wave equation which has the form
$$f_{\omega lm}~~ =~~ (4\pi\omega)^{-1/2}(e^{-i\omega \tV}~ +~ e^{-i\omega {\cal G}(\tU)})
R^{-1} Y_{lm}\eqno(42)$$
in the asymptotic region. These modes reduce to the standard (incoming) modes on $\scrim$ and
have a complicated (outgoing) form on $\scrip$.

The flux of energy radiated to $\scrip$ by the massless scalar field particles is then
given by the expectation value of the off-diagonal component of the stress energy tensor
of the massless scalar field,
$$T^R_T~~ =~~ {1 \over 2} \left\{\phi^{,R}, \phi_{,T}\right\}_+,\eqno(43)$$
where $\{\}_+$ represents the anticommutator. The operator is naturally not well defined but
may be renormalized by point-splitting, giving
$$\langle 0|T^R_T|0\rangle~~ =~~ {1 \over {4\pi R^2}}\sum_{lm} |Y_{lm}|^2 \left[{1 \over 4}
\left({{{\cal G}^{''}}\over {{\cal G}'}}\right)^2~ -~ {1 \over 6} {{{\cal G}^{'''}}\over
{{\cal G}'}}\right] \eqno(44)$$
The total power radiated across a sphere or radius $r$ at late times is therefore
$$P~~ =~~ \int \langle 0 | T^R_T | 0 \rangle R^2 \sin\theta d\theta d\phi~~ =~~ {1 \over
{4\pi}} \sum_{l,m} \left[{1 \over 4} \left({{{\cal G}^{''}}\over {{\cal G}'}}\right)^2~ -~ {1
\over 6} {{{\cal G}^{'''}}\over {{\cal G}'}}\right]\eqno(45)$$
where the sum is over all (angular momentum) modes, $l,m$, and is thus formally infinite. However,
the geometric optics approximation is invalid for the higher angular momentum modes. This is due
to the centrifugal effective potential that causes the mode function to scatter to infinity before
it can pass through the region of high curvature. One expects that this effect will reduce the flux
considerably for modes of large angular momentum, therefore the expression
in (45) is expected to give a good approximation for small $l$ but for large $l$ the radiated power
is expected to diminish rapidly, becoming effectively vanishing\footnote{\dagger}{We wish to
thank S. Fulling for clarifying this point and for correcting certain bibliographical errors in the
original version of this paper.}${}^{[17]}$. One can write the same expression
in terms of ${\cal F}$ as follows
$$P~~ =~~ \int \langle 0 | T^R_T | 0 \rangle R^2 \sin\theta d\theta d\phi~~ =~~ {1 \over
{24\pi}} \sum_{l,m}\left[{{{\cal F}{'''}} \over {({\cal F}')^3}}~ -~ {3 \over 2} \left({{{\cal F}
{''}} \over {{\cal F}^{'2}}} \right)^2\right]\eqno(46)$$
The function ${\cal F}(\tV)$ is the result of
tracing a null ray coming in at $\tV=$ const. from $\scrim$ traveling across the boundary, through
the center and out across the boundary again to become the ray $\tU=$ const.$={\cal F}(\tV)$ on $\scrip$.
The heart of the geometric optics approximation is therefore in the determination of the function
${\cal F}(\tV)$. As we have set it up, this is now an easy task for the problem at hand.

Therefore, consider a ray $\tV=$ const. in the infinite past. We are interested only in the region
on $\scrip$ that is close to the Cauchy horizon, so the approximations in (33) and (38) will suffice.
As the null ray crosses the boundary, we have
$$\tV(v)~~ =~~ \tV^{(0)}~ +~ \Gamma_+ \left(-{v\over {r_o}} \right)^{1 \over {\gamma_+}}
\eqno(47)$$
This expression can be inverted to give
$$v(\tV)~~ =~~ -~ r_o \left[ {{\tV^{0} - \tV} \over {|\Gamma_+|}} \right]^{\gamma_+} \eqno(48)$$
where we have used the fact that $\Gamma_+$ is negative. Next, reflecting about the center (here,
$u=v$) gives
$$u(\tV)~~ =~~ -~ r_o \left[ {{\tV^{0} - \tV} \over {|\Gamma_+|}} \right]^{\gamma_+} \eqno(49)$$
Now as the outgoing ray crosses the outer boundary, we have the relation
$$\eqalign{\tU(u)~~ &=~~ \tU^{(0)}~ -~  \Gamma_- \left(-{u \over {r_o}} \right)^{1 \over {\gamma_-}}
\cr \rightarrow~~ \tU(\tV)~~ &=~~ \tU^{(0)}~ -~ |\Gamma_-|\left[{{\tV^{0} - \tV}\over {|\Gamma_+|}}
\right]^{{\gamma_+} \over {\gamma_-}}\cr} \eqno(50)$$
where now we have used the fact that $\Gamma_-$ is negative. Thus, the right hand side of (50) is
${\cal F}(\tV)$ and it has the form
$${\cal F}(\tV)~~ =~~ A~ -~ B (\tV^{(0)} - \tV)^{{\gamma_+} \over {\gamma_-}} \eqno(51)$$
where $B$ is a positive constant which is given in terms of the roots, $\alpha_\pm$ given before.
We can now write down the power radiated as a function of $\tV$.
>From (46), it follows that
$$P(\tV)~~ =~~ {1 \over {48\pi B^2}} \left[{{1-\gamma^2} \over {\gamma^2(\tV^{(0)} - \tV)^{2\gamma}
}}\right] ~~~~~ \gamma \neq 0 \eqno(52)$$
where
$$\gamma~~  =~~ {{\gamma_+} \over {\gamma_-}}\eqno(53)$$
The expression can also be expressed as a function of the outgoing null coordinate $\tU$ on
$\scrip$ by simply exchanging $\tV$ for $\tU$ in (52), using (50), or directly computing the
radiated power from (45). The result is
$$P(\tU)~~ =~~ {1 \over {48\pi}} \left[{{1-\gamma^2} \over {\gamma^2(\tU^{(0)} - \tU)^{2}
}}\right]\eqno(54)$$
Now, $0 < \gamma \leq 1$  for all $a$ in the range that admits naked singularities (as shown in
figure IV), approaching unity in the limit $a \rightarrow 0$ and decreasing to a minimum of $\sim
0.6$. It is imaginary when $a>a_c$, which is to be expected because there is no outgoing ray
from the origin and the entire treatment breaks down. One sees clearly that the flux diverges as
the Cauchy horizon is approached.
\vskip 0.1in
\centerline{\epsfxsize=2.5in \epsfbox{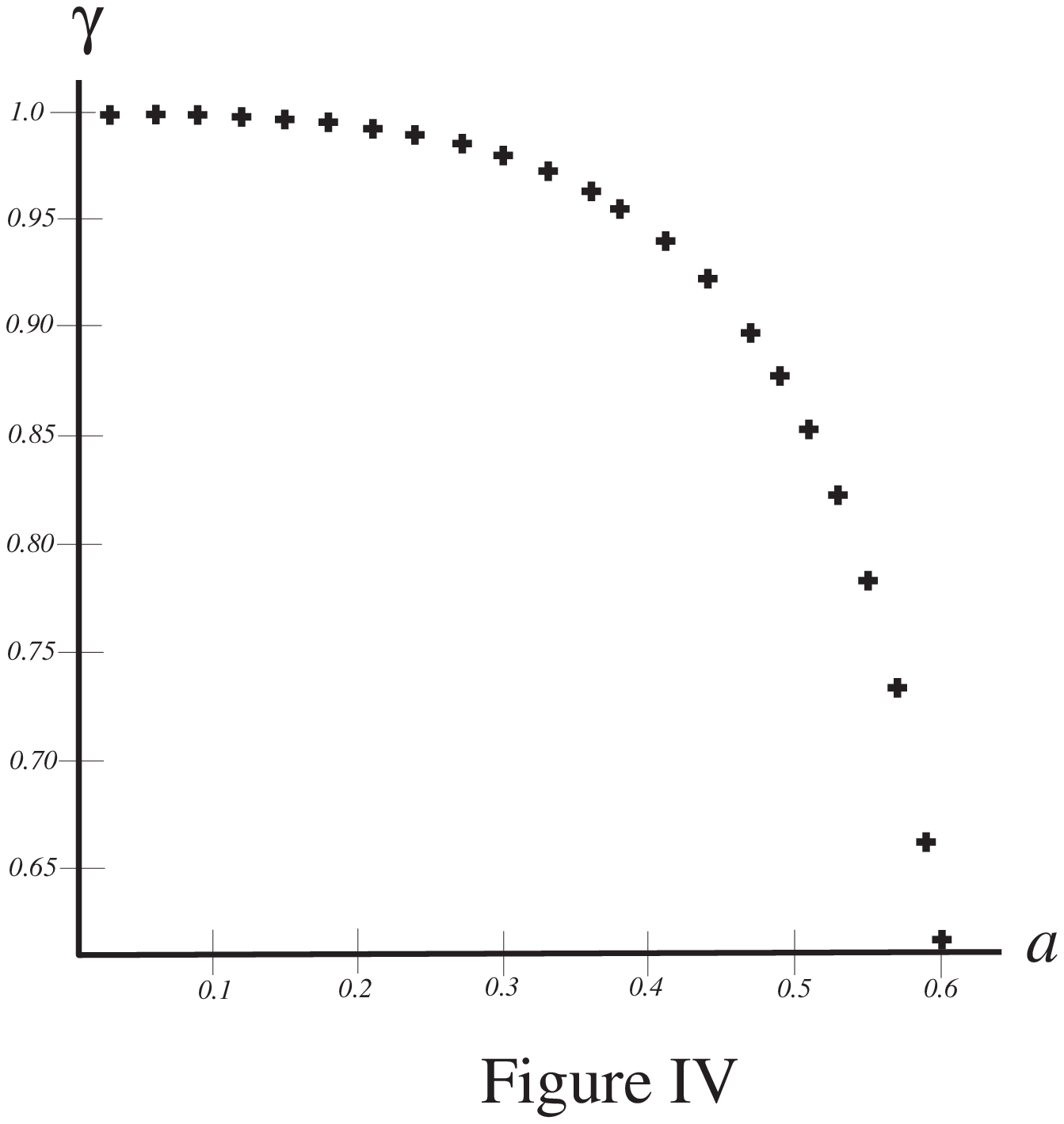}}
\vskip 0.1in

In this article we have examined scalar particle production in the neighborhood of the Cauchy horizon
of the shell focusing naked singularity formed by the self similar collapse of an inhomogeneous,
marginally bound, dust cloud and found that the radiated power will increase rapidly as the Cauchy
horizon is approached. Particle production is a purely kinematical phenomenon and not directly related
to the Einstein equations. The phenomenon we have described should not depend, therefore, on the
particular solution that has been examined. On the contrary, we expect this to occur generically when
regions of high curvature are exposed to the asymptotic observer.

What consequences could this have? We imagine a dust cloud whose initial conditions are such as to
lead to the formation of a naked singularity. As the successive shells begin to form the singularity,
the uncertainty principle takes control leading to a steadily growing and finally intense outgoing
radiation of energy from the cloud as the Cauchy horizon is approached. In the final stages of
collapse, this radiation should occur at extremely high energies that will possibly be visible to
the asymptotic observer and will provide signatures of the behavior of quantum fields in curved
spacetimes. Owing to the strong back reaction of the spacetime in the final stages, we do not expect
that the naked singularity will actually form. Indeed, the CCH may have it's origins in precisely such
an effect.

It may be argued that the correct arena in which such strong gravitational fields and the back reaction
of spacetime should be studied is string theory, as string theory provides a consistent quantum theory
of gravity. The outcome suggested above has been verified in some two dimensional models of string
gravity.${}^{[18]}$ We believe it is of interest to pursue this investigation, both from a theoretical
standpoint as well as an experimental.
\vskip 0.25in

\noindent{\bf Acknowledgements:}

\noindent We acknowledge the partial support of the {\it Junta Nacional de Investiga\c{c}\~ao
Cient\'\i fica e Tecnol\'ogica} (JNICT) Portugal, under contract number CERN/S/FAE/1172/97. C.V.
and L.W. acknowledge the partial support of NATO, under contract number CRG 920096 and L.W.
acknowledges the partial support of the U. S. Department of Energy under contract number
DOE-FG02-84ER40153.
\vskip 0.5in

\noindent{\bf Figure Captions:}

{\item{\bf I.}}Penrose diagram for the spacetime of the marginally bound, self-similar collapse of
inhomogeneous dust. The heavy dashed outgoing null ray is the Cauchy horizon at $y=\alpha_-$ and
the lighter null lines trace a ray that originated at some advanced time $\tV$ on $\scrim$
and crosses $\scrip$ close to the Cauchy horizon.

{\item{\bf II.}}The behavior of $\Gamma_-(\alpha_-)$ as a function of the mass parameter $a$ and for
values of $a$ that admit naked singularity solutions.

{\item{\bf III.}}The behavior of $\Gamma_+(\alpha_+)$ as a function of the mass parameter $a$ and for
values of $a$ that admit naked singularity solutions.

{\item{\bf IV.}}The behavior of the exponent $\gamma=\gamma_+/\gamma_-$ in (53) as a function of
the mass parameter $a$ and for values of $a$ that admit naked singularity solutions.
\vfill\eject

\noindent{\bf References:}

{\item{[1]}}R. Penrose, Riv. Nuovo Cimento {\bf 1} (1969) 252; in {\it General Relativity, An Einstein
Centenary Survey}, ed. S. W. Hawking and W. Israel, Cambridge Univ. Press, Cambridge, London (1979)
581. In its original form, the Cosmic Censorship Hypothesis (CCH) essentially states that: {\it
no physically realistic collapse, evolving from a well posed initial data set and satisfying the
dominant energy condition, results in a singularity in the causal past of null infinity}. There is
also a strong version of the CCH which states that: {\it no physically realistic collapse leads to a
locally timelike singularity}.

{\item{[2]}}P. S. Joshi, {\it Global Aspects in Gravitation and Cosmology}, Clarendon Press, Oxford,
(1993).

{\item{[3]}}J. R. Oppenheimer and H. Snyder (1939) Phys. Rev. {\bf 56} 455

{\item{[4]}}R. C. Tolman (1934) Proc. Nat. Acad. Sci. USA {\bf 20} 169; H. Bondi (1947) Mon. Not. Astron.
Soc. {\bf 107} 410

{\item{[5]}}P. Yodzis, H.-J Seifert and H. M\"uller zum Hagen, Commun. Math. Phys. {\bf 34} (1973) 135;
Commun. Math. Phys. {\bf 37} (1974) 29; D. M. Eardley and L. Smarr, Phys. Rev. D {\bf 19}, (1979) 2239;
D. Christodoulou, Commun. Math. Phys. {\bf 93} (1984) 171; R. P. A. C. Newman, Class. Quantum Grav.
{\bf 3} (1986) 527; B. Waugh and K. Lake, Phys. Rev. D {\bf 38} (1988) 1315; V. Gorini, G. Grillo and
M. Pelizza, Phys. Lett. A {\bf 135} (1989) 154; G. Grillo, Class. Quantum Grav. {\bf 8} (1991) 739; R.
N. Henriksen and K. Patel, Gen. Rel. Gravn. {\bf 23} (1991) 527; I. H. Dwivedi and S. Dixit, Prog.
Theor. Phys. {\bf 85} (1991) 433; P. S. Joshi and I.H. Dwivedi, Phys. Rev {\bf D47} (1993) 5357; I. H.
Dwivedi and P.S. Joshi, Comm. Math. Phys. {\bf 166} (1994) 117; S. Jhingan, P.S. Joshi, T. P. Singh,
Class. Quant. Grav. {\bf 13} (1996) 3057;I. H. Dwivedi, P. S. Joshi, Class. Quant. Grav. {\bf 14}
(1997) 1223.

{\item{[6]}}T. P. Singh and P. S. Joshi, Class. Quant. Grav. {\bf 13} (1996) 559.

{\item{[7]}}W. A. Hiscock, L. G. Williams and D. M. Eardley, Phys. Rev. {\bf D26} (1982) 751; Y. Kuroda,
Prog. Theor. Phys. {\bf 72} (1984) 63; A. Papapetrou, in {\it A Random Walk in General Relativity},
Eds. N. Dadhich, J. K. Rao, J. V. Narlikar and C. V. Vishveshwara, Wiley Eastern, New Delhi (1985);
G. P. Hollier, Class. Quantum Grav. {\bf 3} (1986) L111; W. Israel Can. Jour. Phys. {\bf 64} (1986)
120; K. Rajagopal and K. Lake, Phys. Rev. {\bf D35} (1987) 1531; J. Lemos, Phys. Rev. Lett. {\bf 68}
(1992) 1447.

{\item{[8]}}A. Ori and T. Piran, Phys. Rev. {\bf D42} (1990) 1068.

{\item{[9]}}D. Christodoulou, Commun. Math. Phys. {\bf 105} (1986) 337; {\it ibid.} {\bf 106} (1986) 587;
{\it ibid.} {\bf 109} (1987) 591; {\it ibid.} {\bf 109} (1987) 613; Commun. Pure Appl. Math. {\bf  XLIV}
(1991) 339; {\it ibid.} {\bf XLVI} (1993) 1131; Ann. Math. {\bf 140} (1994) 607; M. W. Choptuik, Phys. Rev.
Lett. {\bf 70} (1993) 9; C. Gundlach, J. Pullin and R. Price, Phys. Rev. {\bf D49} (1994) 890; D.
Garfinkle, Phys. Rev. {\bf D51} (1995) 5558; R. S. Hamade and J. M. Stewart, Class. and Quantum Grav.
{\bf 13} (1996) 497; M. D. Roberts, Gen. Rel. Grav. {\bf 21} (1989) 907; V. Husain, E. Martinez and D.
Nunez, Phys. Rev. {\bf D50} (1994) 3783; J. Traschen, Phys. Rev. {\bf D50} (1994) 7144; Y. Oshiro, K.
Nakamura and A. Tomimatsu, Prog. Theor. Phys. {\bf 91} (1994) 1265; P. R. Brady, Class. Quantum Grav.
{\bf 11} (1994) 1255; Phys. Rev. {\bf D51} (1995) 4168.

{\item{[10]}}C. Gundlach, Phys. Rev. Lett. {\bf 75} (1995) 3214; Phys. Rev. {\bf D55} (1997) 695; A.
M. Abrahams and C. R. Evans, Phys. Rev. Lett. {\bf 70 } (1993) 2980; C. R. Evans and J. S. Coleman,
Phys. Rev. Lett. {\bf 72} (1994) 1782; D. Maison, Phys. Letts. {\bf B366} (1996) 82; E. W. Hirschmann
and D. M. Eardley, Phys. Rev.{\bf D51} (1995) 4198; Phys. Rev {\bf D52} (1995) 5850; Phys. Rev. {\bf D56}
(1997) 4696; D. M. Eardley, E. W. Hirschmann and J. H. Horne, Phys. Rev. {\bf D52} (1995) 5397; R. S.
Hamade, J. H. Horne and J. M. Stewart, Class. Quant. Grav. {\bf 13} (1996) 2241.

{\item{[11]}}P.S. Joshi and T.P. Singh, Phys. Rev. {\bf 51} (1995) 6778, and references therein.

{\item{[12]}}G. T. Moore, J. Math. Phys. {\bf 9} (1979) 2679.

{\item{[13]}}S. W. Hawking, Comm. Math. Phys. {\bf 43} (1975) 199.

{\item{[14]}}B. S. DeWitt, Phys. Rep. {\bf 19C} (1975) 295.

{\item{[15]}}S. A. Fulling and P.C.W. Davies, Proc. R. Soc. London {\bf A348} (1976) 393; {\it ibid.}
{\bf A356} (1977) 237. See also, N. D. Birrel and P.C.W. Davies, {\it Quantum Fields in
Curved Space}, Cambridge Monographs in Math. Phys., Cambridge University Press, London (1982).

{\item{[16]}}L. H. Ford and Leonard Parker, Phys. Rev. {\bf D17} (1978) 1485.

{\item{[17]}}This effect is discussed in ref.[14]. See also S. M. Christensen and S. A. Fulling,
Phys. Rev. {\bf D15} (1977) 2088.

{\item{[18]}}Cenalo Vaz and Louis Witten, Phys. Letts. {\bf B325} (1994) 27; Class. Quant.
Grav. {\bf 12} (1995) 1; {\it ibid.} {\bf 13} (1996) L59; Nucl. Phys. {\bf B487} (1997)
409.
\bye